\documentclass[%
 preprint, 
 amsmath,amssymb,
 aps, physrev,
]{revtex4-2}

\usepackage{graphicx}
\usepackage{subcaption}
\usepackage{algorithm}

\usepackage[colorlinks=true, linkcolor=blue, citecolor=blue, urlcolor=blue]{hyperref}

\begin{document}

\preprint{APS/123-QED}

\title{\textbf{Nonreciprocal quantum coherence in cavity magnomechanics via the Barnett effect} 
}% 

\author{Jinhao Jia}
 \altaffiliation[School of Physics and Astronomy, Beijing Normal University, Beijing, 100875, China ]{Applied Optics Beijing Area Major Laboratory, Beijing Normal University, Beijing, 100875, China}
 
\author{Yingru Li}
 \altaffiliation[School of Physics and Astronomy, Beijing Normal University, Beijing, 100875, China ]{Applied Optics Beijing Area Major Laboratory, Beijing Normal University, Beijing, 100875, China}

 \author{Juan Huang}
 \altaffiliation[School of Physics and Astronomy, Beijing Normal University, Beijing, 100875, China ]{Applied Optics Beijing Area Major Laboratory, Beijing Normal University, Beijing, 100875, China}
 
\author{Mei Zhang}%
 \email{Contact author: zhangmei@bnu.edu.cn}
\affiliation{School of Physics and Astronomy, Beijing Normal University, Beijing, 100875, China
}%

% \author{Jinhao Jia}
 
% \affiliation{
%  School of Physics and Astronomy, Beijing Normal University, Beijing, 100875, China
% }%
% \affiliation{
%  Applied Optics Beijing Area Major Laboratory, Beijing Normal University, Beijing, 100875, China
% \affiliation{%
% }%

% \collaboration{CLEO Collaboration}%\noaffiliation

\date{\today}% It is always \today, today,

\begin{abstract}
We theoretically investigate the quantum coherence ans its nonreciprocity in a cavity magnomechanical (CMM) syetem, which consists of a rotating yittrium iron garnet (YIG) sphere and a microwave cavity. By adjusting the direction of the magnetic field, the frequency shift of a magnon mode can be tuned from positive to negative due to the Barnett effect. This effect leads to a significant difference in the system stability and is responsible for the nonreciprocal quantum coherence. We examine how the input power, magnomechanical and magnon-photon coupling rates, decay rates of both the cavity photon modes and the magnon modes influence the quantum coherence. Through careful tuning of system parameters, nearly perfect nonreciprocity can be achieved. Our results provide a controllable mechanism for direction-dependent quantum coherence, with potential applications in nonreciprocal quantum devices and information processing.
\end{abstract}

%main conclusions). Our paper paves the way for highly tunable information processing by modulating the system parameters in the CMM systems via the Barnett effect.

\maketitle

\section{\label{sec:level1}Introduction:}

Quantum coherence describes the correlation between quantum fluctuations of coupled quantum systems \cite{PhysRevLett.119.140402,PhysRevLett.113.140401,PhysRevLett.115.020403,PhysRevA.96.032316,PhysRevA.93.032111,PhysRevLett.116.120404}. It is the characteristic of quantum physics and the fundamental manifestation of quantum superposition principle \cite{PhysRevLett.117.030401,https://doi.org/10.1002/qute.202100040}, which serves to promote a deep understanding of diverse quantum phenomena. Having a long and rich history, the phenomenon of quantum coherence manifests itself across a wide range of physical scales, from macroscopic effects like superfluidity, superradiance, and superconductivity, to microscopic systems such as atoms and qubits \cite{PhysRevLett.113.170401}. Quantum coherence serves as vital resource in various fields of quantum optics \cite{PhysRevA.89.052302,PhysRevLett.94.173602,PhysRevA.90.033812},quantum metrology \cite{PhysRevLett.113.250801}, quantum information processing \cite{RevModPhys.89.041003,2017Faithful} and quantum biology \cite{2007Evidence}. It is worth mentioning that various methods have been proposed to quantify quantum coherence, such as relative entropy of coherence \cite{PhysRevA.93.032111}, geometric and concurrence measurements of coherence \cite{PhysRevLett.115.020403}. Quantitative description of quantum coherence helps to investigate the boundaries between quantum and classical worlds and manipulate quantum correlation between quantum systems \cite{PhysRevLett.117.030801, 201400150}. Recently, macroscopic quantum coherence have been studied in various physical systems, including optomechanical systems \cite{2019Quantum,PhysRevA.94.052314,PhysRevA.96.063819}, Josephson junctions \cite{PhysRevA.97.042103,PhysRevLett.105.177001} and more.

Among these, cavity magnomechanical (CMM) systems have emerged as promising candidates\cite{PhysRevLett.121.203601,doi:10.1126/sciadv.1501286}. Serval theoretical schemes have been proposed to manipulate quantum coherence \cite{PhysRevB.109.064412,PhysRevA.105.063718,article} and achieve nonreciprocal quantum coherence \cite{PhysRevA.109.013719}. CMM systems consist of a ferrimagnetic crystal (typically a yttrium iron garnet sphere) embedded in a microwave cavity. The system supports three modes: a magnon mode (collective spin excitations), a phonon mode (mechanical vibrations) and a photon mode (microwave cavity photons) \cite{PhysRevA.99.021801}. Through magnetostrictive interactions, coherent coupling between the magnon mode and the phonon mode is realized, while the photon modes couples to the magnon mode via magnetic dipole interaction. Compared with conventional optomechanical systems, CMM systems offer several distinct advantages. They exhibit high tunability, as the magnon mode frequency can be precisely controlled by adjusting the strength of an external magnetic field, allowing flexible manipulation of coherent coupling. In addition, due to the high spin density of magnetic materials and low damping rates, CMM systems feature strong light-matter interactions. Building upon these properties, CMM platforms have enabled the exploration of a wide range of quantum and nonlinear phenomena. Theoretical studies have investigated their potential for entangled and squeezed states \cite{PhysRevA.99.021801,PhysRevB.101.014416,PhysRevB.100.174407,PhysRevA.109.013704}, nonreciprocal entanglement\cite{PhysRevA.109.043512,PhysRevA.111.013713,PhysRevApplied.21.034061}, stationary one-way quantum steering \cite{https://doi.org/10.1002/qute.202300374,Zhang:22}, the steady Bell state generation \cite{PhysRevLett.124.053602} and nonreciprocal high-order sidebands \cite{PhysRevA.104.033708}. Recent experimental advances have further demonstrated the versatility of CMM platforms, which have been used to realize magnomechanically induced transparency and absorption \cite{doi:10.1126/sciadv.1501286}, mechanical bistability \cite{PhysRevLett.129.123601}, and coherent microwave-to-optical conversion \cite{PhysRevLett.129.243601}.

Nonreciprocity, resulting from the violation of Lorentz reciprocity \cite{PhysRevApplied.10.047001,RJPotton_2004}, is characterized by an asymmetric transmission of signals in opposite directions. It is crucial for applications in nanophotonic communication systems, nosie-free information processing \cite{2017Non}, and rotating resonators \cite{PhysRevApplied.7.064014,PhysRevApplied.12.034001,PhysRevLett.121.153601,Li:19}. Traditionally, nonreciprocity has been achieved using magneto-optical effects, such as Faraday rotation, but such devices are typically bulky and unsuitable for on-chip integration \cite{PhysRevLett.100.013904,PhysRevLett.105.233904}. This has motivated the search for alternative microscale mechanisms to induce nonreciprocity. 
 The Barnett effect, in which the mechanical rotation of a magnetic body induces magnetization through conservation of angular momentum, has recently been observed in both ferromagnetic insulators \cite{PhysRevLett.129.257201,PhysRevB.92.174424,PhysRev.6.239,doi:10.7566/JPSJ.86.011011,10.1063/1.3232221} and nuclear spin systems \cite{PhysRevLett.122.177202,Chudo_2014}. In CMM systems, the rotation of a YIG sphere gives rise to a Barnett-induced frequency shift of the magnon mode. This shift can be tuned by reversing either the direction of rotation or the orientation of the external bias magnetic field \cite{PhysRevLett.129.257201,PhysRevB.92.174424,PhysRev.6.239,doi:10.7566/JPSJ.86.011011}. As a result, the system enables nonreciprocal entanglement generation. In 2025, Tian-Xiang Lu et al proposed a scheme that can achieve nonreciprocal entanglement using the Barnett effect \cite{PhysRevA.111.013713}. In contrast to existing approaches that achieve nonreciprocal entanglement in CMM systems via the magnon Kerr effect \cite{PhysRevA.109.043512} or chiral coupling \cite{FAN2025}, their method does not rely on any topological or chiral structures, nor on Kerr-type nonlinearities. This makes it more accessible and feasible under current experimental conditions.

Inspired by previous studies, we theoretically explore the properties of quantum coherence and its nonreciprocal behavior in the CMM system based on the Barnett effect, which consists of the rotating YIG sphere coupled to the microwave cavity. In this work, we focus on quantifying coherence in bosonic Gaussian states of infinite-dimensional systems, using the coherence measurement based on relative entropy \cite{PhysRevA.93.032111}. Our analysis reveals that the Barnett effect in the microwave cavity induces a notable asymmetry in system stability and quantum coherence when the bias magnetic field is driven from opposite directions. Furthermore, within experimentally feasible parameter regimes, we examine how the driving power, magnomechanical and magnon-photon coupling rates, decay rates and detunings of both the cavity photon modes and the magnon modes influence the quantum coherence. Notably, by appropriately tuning the system parameters and operating within the stable regime, the coherence of the mechanical mode can surpass that of both the microwave photon and magnon modes. This phenomenon can be interpreted as a transfer of quantum coherence from the photon and magnon modes to the mechanical mode. These findings pave the way for the development of novel microscale magnonic structures with promising applications in tunable quantum information processing and communication technologies.

\section{\label{sec:level2}System and dynamics:}
\begin{figure}[htbp]
  \begin{center}
  \begin{subfigure}[b]{0.5\linewidth}\label{figure1a}
    \centering
    \includegraphics[width=\linewidth]{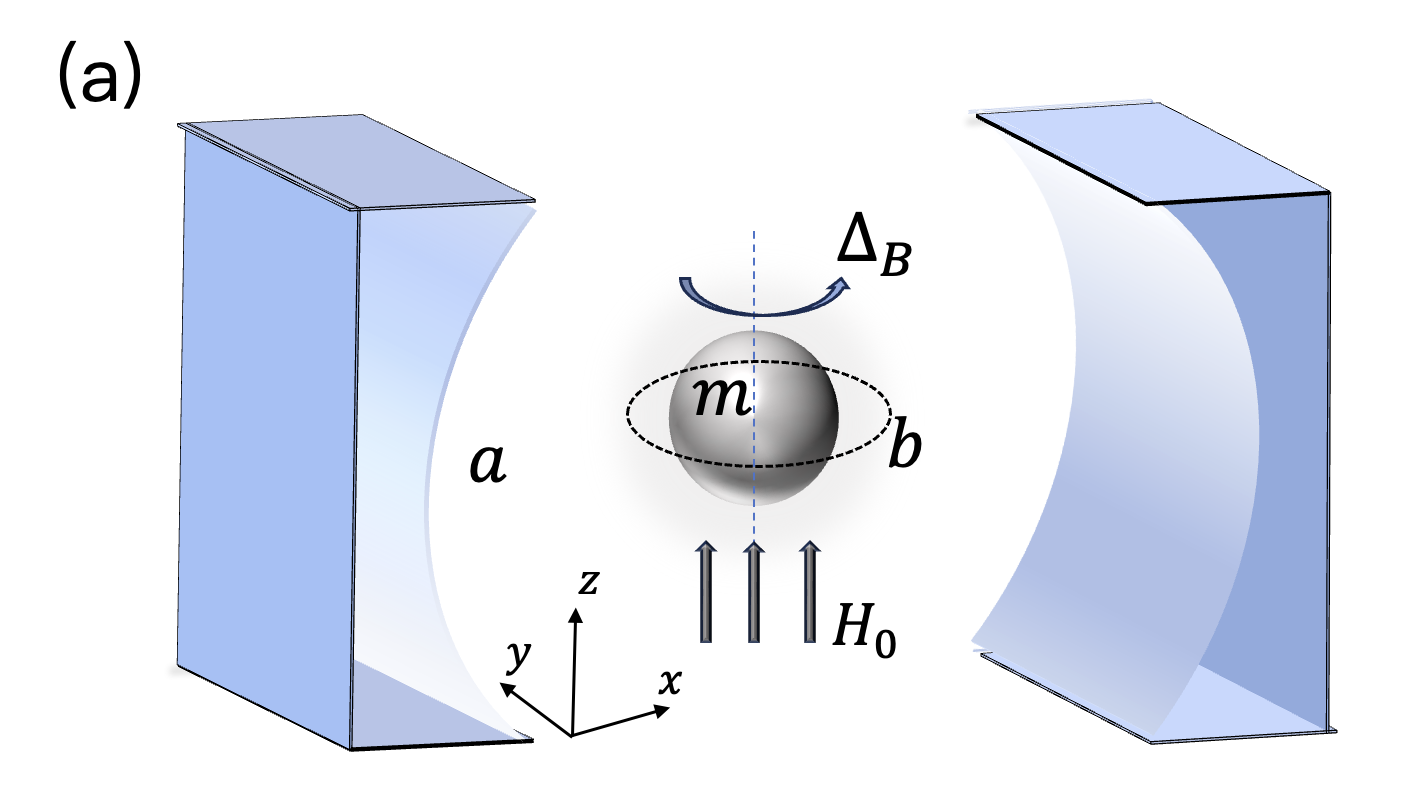} 
 
  \end{subfigure}
  \hfill
  \begin{subfigure}[b]{0.5\linewidth}
    \centering
    \includegraphics[width=\linewidth]{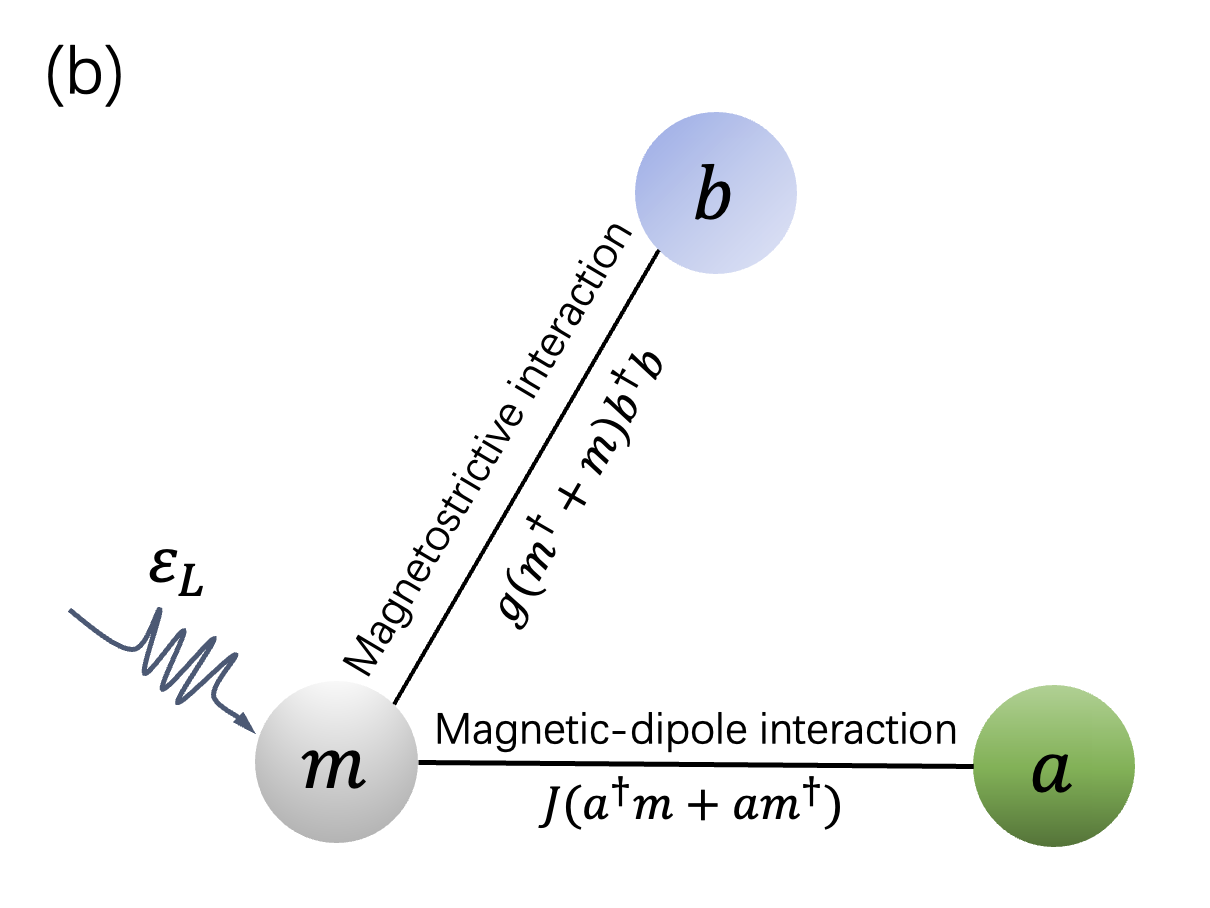}
    
  \end{subfigure}
  \caption{(a) Schematic diagram of the proposed CMM system. It consisted of a YIG sphere and a microwave cavity. The YIG sphere with angular frequency $\Delta_B $ is subjected uniform bias magnetic field $ H_0 $.(b) Sketch of the interactions in the CMM system. The magnon couples to the photon mode via magnetic-dipole interaction, and to the phonon mode through magnetostrictive interaction.}\label{fig:total}
   \end{center}
  \end{figure}

As shown in Fig.~\ref{fig:total}(a), we consider a general model of the CMM system, which consists of a microwave cavity mode, a magnon mode, and a phonon mode \cite{PhysRevLett.121.203601}. The magnon mode, representing collective spin excitations in a YIG sphere, has a frequency $\omega_{m}$ that depends linearly on the applied bias magnetic field $H_{0}$ via $\omega_{m} = \gamma H_{0}$, where $\gamma$ is the gyromagnetic ratio \cite{doi:10.1126/sciadv.1501286}. The magnon mode couples to the phonon mode, with frequency $\omega_{b}$, through magnetostrictive interaction. Simultaneously, the microwave cavity mode, with frequency $\omega_{a}$, couples to the magnon mode via the magnetic-dipole interaction \cite{PhysRevA.99.021801}see Fig.~1(b). When the YIG sphere rotates about the $z$-axis with angular frequency $\Delta_B$, it generates an emergent magnetic field due to the Barnett effect[], expressed as $H_B=\Delta_B/\gamma$ \cite{PhysRevLett.129.257201,PhysRevB.92.174424,PhysRev.6.239,doi:10.7566/JPSJ.86.011011,10.1063/1.3232221}. Experimental results have confirmed that the induced magnetization is directly proportional to $\Delta_B$, and its polarity changes with the direction of rotation \cite{PhysRevB.92.174424}. Due to angular momentum conservation, the magnon frequency shifts from $\omega_{m}$ to $\omega_{m} + \Delta_B$. By keeping the YIG sphere rotating counterclockwise at a fixed angular frequency $\Delta_B$ and reversing the direction of the external magnetic field (along $+z$ or $-z$), one can tune the Barnett-induced frequency shift from positive to negative values.

The Hamiltonian of the CMM systems in a frame rotating with the drive frequency reads \cite{PhysRevA.104.033708}

\begin{align}
{H}/\hbar &= \Delta_{a} a^{\dagger} a 
+ (\Delta_{m}+\Delta_{B} )m^{\dagger} m+\omega_b b^{\dagger} b+J(a^\dagger m + a m^{\dagger})  
 + g m^{\dagger} m (b^{\dagger} + b)+\mathrm{i}\varepsilon_{l}(m^{\dagger}+m)
\end{align}
where $\Delta_{a(m)}=\omega_{a(m)}-\omega_l$. $o(o^{\dagger})(o=a,m,b)$ are the annihilation (creation) operators of the microwave cavity mode, the magnon mode and the phonon mode. They satisfy the commutation relation $[o,o^{\dagger}]=1$. $J$ is the magnetic dipole interaction strength constants. The bare magnon-mechanical coupling strength, $g
$, is generally weak but can be effectively enhanced by applying strong driving fields to the YIG sphere \cite{0Entanglement}. $\varepsilon_l=\sqrt{2\kappa_mP/\hbar\omega_l}$ is the amplitude of the driving fields, where $P_m$ is the driving power and $\omega_l$ the driving frequency.

The Heisenberg-Langevin equation describing the dynamics are given by
\begin{eqnarray}
	\dot{a}&=&-(\mathrm{i}\Delta_{ a}+\kappa_{a})a-\mathrm{i}Jm+\sqrt{2\kappa_{a}}{a}^{\textrm{in}}(t),\\ \nonumber
    \dot{m}&=&-(\mathrm{i}\Delta_{m}+\mathrm{i}\Delta_{B}+\kappa_{m})m-\mathrm{i}J a-\mathrm{i}g(b^{\dagger}+b)+\varepsilon_{l}+\sqrt{2\kappa_{m}}{m}^{\textrm{in}}(t)\\\nonumber
	\dot{b}&=&-(\mathrm{i}\omega_{ b}+\kappa_{b})b-\mathrm{i}g m^{\dagger}m +\sqrt{2\kappa_{b}}{b}^{\textrm{in}}(t), 
\end{eqnarray}
where $\kappa_{o}$ are the decay rates of the microwave cavity mode, the magnon mode and the phonon mode respectively. The operator $\hat{o}_{\text{in}}(t)$ represents the input noise associated with the target mode $o$. The input noise has zero mean and is characterized by the following correlation functions: 
\begin{eqnarray}
\langle o^{\textrm{in}}(t)o^{\textrm{in}\dagger}(t^{\prime})&=&[N_{o}(\omega_{o})+1]\delta(t-t^{\prime}), \\ \nonumber
\langle o^{\textrm{in}\dagger}(t)o^{\textrm{in}}(t^{\prime})\rangle&=&N_{o}(\omega_{0})\delta(t-t^{\prime}),
\end{eqnarray}
where $N_{0}(\omega_{0})=[\exp(\frac{\hbar\omega_{0}}{k_\textrm{B}T})-1]^{-1}$is the mean thermal excitation number, $k_\textrm{B}$ the Boltzmann constant, T the environment temperature.

To linearize the system dynamics, we decompose each operator into its steady-state mean value and a small fluctuation: $o = \langle o \rangle + \delta o$. By substituting the expression into the quantum Langevin equations (QLEs) and neglecting second-order fluctuation terms, we derive the linearized dynamical equations governing the fluctuations. From these equations, the steady-state solution for each operator can be obtained.
\begin{eqnarray}
a_s &=& \frac{-\mathrm{i}Jm_s}{\mathrm{i}\Delta_a+\kappa_a},\\ \nonumber
m_s &=& \frac{\varepsilon_{l}(\mathrm{i}\Delta_a+\kappa_a)}{J^2 + (\mathrm{i}\Delta_a+\kappa_a)(\mathrm{i}\tilde\Delta_{m}+\mathrm{i}\Delta_B+\kappa_m)},\\ \nonumber
b_s &=&\frac{-\mathrm{i}g \mid m_s\mid^2}{\mathrm{i}\omega_b+\kappa_b},\label{eq4}
\end{eqnarray}
where $\tilde\Delta_{m}=\Delta_m+g(b_s+b_s^{*}).$ Then we introduce the quadrature fluctuation operators defined as $X_{0(o^{\mathrm{in}})}(t)=\frac{1}{\sqrt{2}}(o(o^{\mathrm{in}})+o^{\dagger}(o^{\mathrm{in}\dagger}))$ and $Y_{o(o^{\mathrm{in}})}(t)=\frac{1}{\sqrt{2}\mathrm{i}}(o(o^{\mathrm{in}})-o^{\dagger}(o^{\mathrm{in}\dagger})$. The linearized QLEs describing the quadrature fluctuations can be written in matrix form $\dot{u}(t)=Af(t)+n(t)$, where $u^{T}(t)=[\delta X_{a}(t),\delta Y_{a}(t),\delta X_m(t),\delta Y_m(t), \delta X_{b}(t), \delta Y_{b}(t)]$ and~$n^{T}(t)=[\sqrt{2\kappa_{a}}X_{a_{}}^{\mathrm{in}},\sqrt{2\kappa_{a_{}}}Y_{a_{}}^{\mathrm{in}},\sqrt{2\kappa_{m_{}}}X_{m_{}}^{\mathrm{in}}, \sqrt{2\kappa_{m_{}}}Y_{m_{}}^{\mathrm{in}},\sqrt{2\kappa_{b_{}}}X_{b_{}}^{\mathrm{in}},\sqrt{2\kappa_{b_{}}}Y_{b_{}}^{\mathrm{in}}]$. The drift matrix $A$ is given as follows,
\begin{equation}
A = \left( \begin{array}{c c c c c c c}
    -\kappa_a & \Delta_a & 0 & 0 & 0 & J   \\
    -\Delta_a & -\kappa_a & 0 & 0 & -J & 0   \\
    0&J & -g_2 & 0 & -\kappa_m & \tilde\Delta^{\prime}_{m}  \\
    -J& 0 & g_1&0 & -\tilde\Delta^{\prime}_{m} &  -\kappa_m \\
    0& 0 & -\kappa_b & \omega_b & 0 & 0   \\
    0 & 0 & \omega_b & -\kappa_b & g_1 & g_2   \\
\end{array} \right),
\end{equation}
where $g_1=-g(m_s+m_s^{*})$,$g_2=\mathrm{i}(m_s-m_s^{*})$, $\tilde\Delta^{\prime}_{m}=\tilde\Delta_{m}+\Delta_B$.

The steady-state behavior of quantum fluctuations can be fully characterized by the covariance matrix $\mathbf{V}$, a $6 \times 6$ real symmetric matrix with elements defined as $V_{ij} = \frac{1}{2} \langle f_i(\infty) f_j(\infty) + f_j(\infty) f_i(\infty) \rangle$, where $\mathbf{f}_{i(j)}(t)$ denotes the vector of fluctuation operators ($i, j = 1, 2, \dots, 6$). In the limit $t \to \infty$, the covariance matrix $\mathbf{V}$ satisfies the Lyapunov equation \cite{PhysRevA.105.063718},
\begin{equation}
\mathbf{A} \mathbf{V} + \mathbf{V} \mathbf{A}^\mathrm{T} = -\mathbf{D}, \label{eq:lyapunov}
\end{equation}
where $\mathbf{D}$ is the diffusion matrix given by
$D=\mathrm{diag}[\kappa_{a}(2N_{a}+1),\kappa_{a}(2N_{a}+1),\kappa_{m}(2N_{m}+1),\kappa_{m}(2N_{m}+1),\kappa_{b}(2N_{b}+1),\kappa_{b}(2N_{b}+1)]$.
The coupled system reaches a steady state when the Routh-Hurwitz criterion is satisfied, which requires adjusting parameters to keep all eigenvalues of matrix A with negative real parts \cite{PhysRevA.35.5288}. In this work, we focus on the numerical evaluation of quantum properties in the CMM system through the covariance matrix $V^{\text{tot}}$.

Then we evaluate the quantum coherence of the target modes within the coupled system. In the context of continuous-variable systems, quantum coherence characterizes the correlations between quantum fluctuations of quadrature operators and is essential for the generation of entanglement. In this work, we quantify coherence using the relative entropy for Gaussian states, which depends on both the first and second statistical moments of the state \cite{PhysRevA.93.032111}. Consider a single-mode Gaussian state $\rho(\mathbf{V}_o, \vec{d}_o)$, where $\mathbf{V}_o$ denotes the covariance matrix and $\vec{d}_o = (d_{o1}, d_{o2})$ is defined by the mean values of the quadrature fluctuations, with $d_{o1} = (o_s+o_s^{*})/\sqrt{2}$ and $d_{o2} = (m_s-m_s^{*}/)\sqrt{2}\mathrm{i}$. The covariance matrix $\mathbf{V}_o$ is extracted from the full covariance matrix $\mathbf{V}$.

For the photon, the magnon, and the phonon modes, the respective covariance matrices are given by
\begin{equation}
\mathbf{V}_a = 
\begin{pmatrix}
V_{11} & V_{12} \\
V_{21} & V_{22}
\end{pmatrix}, \quad
\mathbf{V}_m = 
\begin{pmatrix}
V_{33} & V_{34} \\
V_{43} & V_{44}
\end{pmatrix}, \quad
\mathbf{V}_b = 
\begin{pmatrix}
V_{55} & V_{56} \\
V_{65} & V_{66}
\end{pmatrix}.
\end{equation}

The quantum coherence of a target mode $o$ can be quantified as
\begin{equation}
C_o[\rho(\mathbf{V}_o, \vec{d}_o)] = -F(\nu_o) + F(2\bar{n}_o + 1),
\label{eq:coherence_criterion}
\end{equation}
where $F(X) = \frac{X+1}{2}\log_2\left(\frac{X+1}{2}\right) - \frac{X-1}{2}\log_2\left(\frac{X-1}{2}\right)$, and $\nu_o = \sqrt{\det(\mathbf{V}_o)}$ is the symplectic eigenvalue of mode $o$. The effective thermal occupation $\bar{n}_o$ is defined as $\bar{n}_o = \left[ \mathrm{Tr}(\mathbf{V}_o) + d_{o1}^2 + d_{o2}^2 - 2 \right]/4$. The total quantum coherence of this three-mode coupled system can be evaluated in a similar way as follows \cite{article},
\begin{equation}
C_{\text{tot}} = \mathcal{C}(V^{\text{tot}}) = \sum_{i=1}^{3} F(2n_{i,\text{av}} + 1) - F(\nu_i),
\end{equation}
the values \(\nu_i\) are elements of the symplectic spectrum \(\nu\) of the total covariance matrix \(V^{\text{tot}}\), which can be obtained from the eigenvalues of \(|i\tilde{\Omega} V^{\text{tot}}|\). Here, \(\tilde{\Omega}\) is a \(6 \times 6\) symplectic matrix defined as:

\begin{equation}
\tilde{\Omega} = \bigoplus_{i=1}^{3} \varpi_i, \quad \varpi_i \equiv 
\begin{bmatrix}
0 & 1 \\
-1 & 0
\end{bmatrix}.
\end{equation}

\section{\label{sec:level3}Numerical simulations and discussions:}
In this section, we numerically investigate the nonreciprocal quantum coherence induced by the Barnett effect in the CMM system using experimentally feasible parameters. Specifically, we focus on the nonreciprocal behavior of the single-mode coherences $C_a$, $C_b$, and $C_m$, and analyze how different physical factors such as the driving power, the magnomechanical and magnon-photon coupling rates, decay rates affect the nonreciprocity. We also examine the impact of these parameters on the total coherence $C_t$, providing a comprehensive understanding of the tunability and control of coherence in such hybrid systems. The parameters we choose are $\omega_a/2\pi=10~\mathrm{GHz}$, $\omega_b/2\pi=20~\mathrm{MHz}$, $\omega_l/2\pi=10~\mathrm{GHz}$, $\kappa_a/2\pi= 1~\mathrm{MHz}$, $\kappa_m/2\pi= 1~\mathrm{MHz}$, $\kappa_b/2\pi=100~\mathrm{Hz}$, $\tilde{\Delta}_{m}=0.9\omega_b$, $J/2\pi=1~\mathrm{MHz}$, $g/2\pi=0.1~\mathrm{Hz}$ and $T=10~\mathrm{K}$ \cite{PhysRevLett.121.203601,doi:10.1126/sciadv.1501286}. In the process of numerical simulation, the steady state is ensured by the Routh-Hurwitz criterion, which requires all eigenvalues of the drift matrix to have negative real parts. 
 
\begin{figure}[H]

    \includegraphics[width=\linewidth]{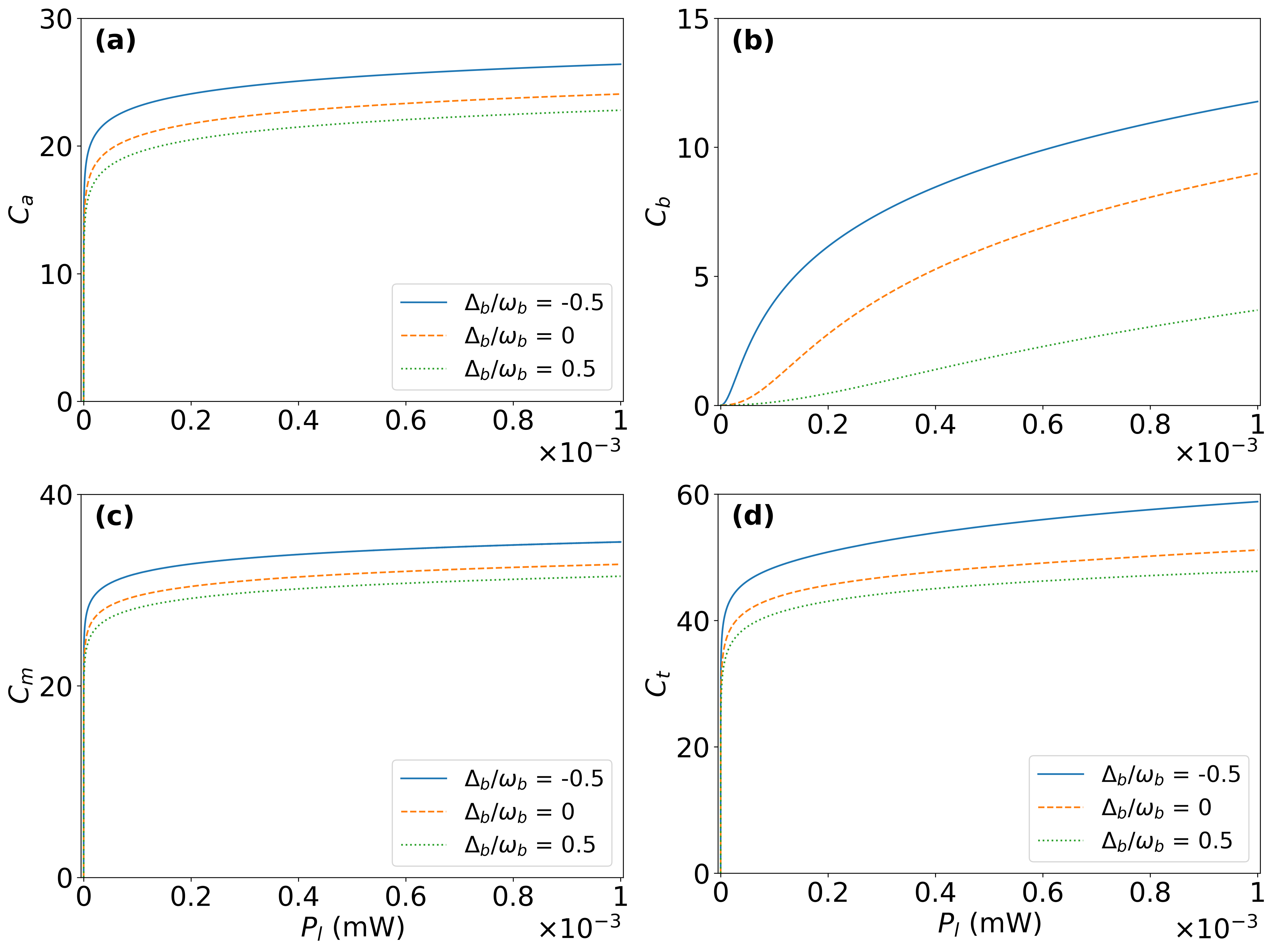}
    \centering
    \caption{Quantum coherences $C_a$,$C_b$,$C_m$,$C_t$ as functions of the driving power $P$. All parameters are in the main text.}
    \label{fig2}
\end{figure}

Figures \ref{fig2}(a)–\ref{fig2}(d) display the behavior of the single-mode quantum coherences $C_a$, $C_b$, $C_m$ and the three-mode quantum coherence as functions of the driving power. It is observed that all quantum coherences emerge rapidly as the driving power increases from zero to a finite value. This behavior can be attributed to the physical mechanism underlying the dynamics of the system. As indicated in Eq.~\ref{eq4}, the steady-state solutions of all modes vanish in the absence of external driving, resulting in zero quantum coherence. To investigate how various parameters influence the emergence of nonreciprocal quantum coherence, we focus on regimes with moderate driving power in the subsequent numerical simulation process. The trends of all quantum coherences with respect to driving power are qualitatively similar, which stems from a common physical origin: the enhancement of the magnomechanical interaction suppresses environmental decoherence in both the photon mode and the magnon modes, facilitating the buildup of quantum coherence. Moreover, the direction of the applied magnetic field significantly influences the coherence behavior. Specifically, relative to the stationary case ($\Delta_B = 0$), the quantum coherence is enhanced for $\Delta_B < 0$, whereas it is suppressed for $\Delta_B > 0$. These observations demonstrate the emergence of nonreciprocal quantum coherence, which is inherently constrained by the stability of the system. When the rotation of the YIG sphere is fixed, reversing the direction of the magnetic field leads to nonreciprocal stability in the CMM system, thereby inducing direction-dependent quantum coherence. Furthermore, we observe that the quantum coherences associated with the the photon and the magnon modes consistently exceed that of the phonon mode, i.e., \( C_a > C_b \) and \( C_m > C_b \). This behavior can be attributed to the typically stronger coupling of the phonon mode to its thermal environment, which induces greater decoherence compared to the photon and magnon modes \cite{PhysRevA.109.013719}.

Next, we investigate the effect of the coupling rate J and the angular frequency $\Delta_{B}$ on the single-mode quantum coherence $C_o$ and the total coherence $C_t$. To precisely characterize the nonreciprocal quantum coherence, we propose the bidirectional contrast ratio $I$ to access the nonreciprocity, defined as follows \cite{PhysRevA.111.013713}:

\begin{align}
I=\frac{\mid C_o(\Delta_B>0)-C_o(\Delta_B<0)\mid}{C_o(\Delta_B>0)+C_o(\Delta_B<0)}
\end{align}

\begin{figure}[H]

    \includegraphics[width=\linewidth]{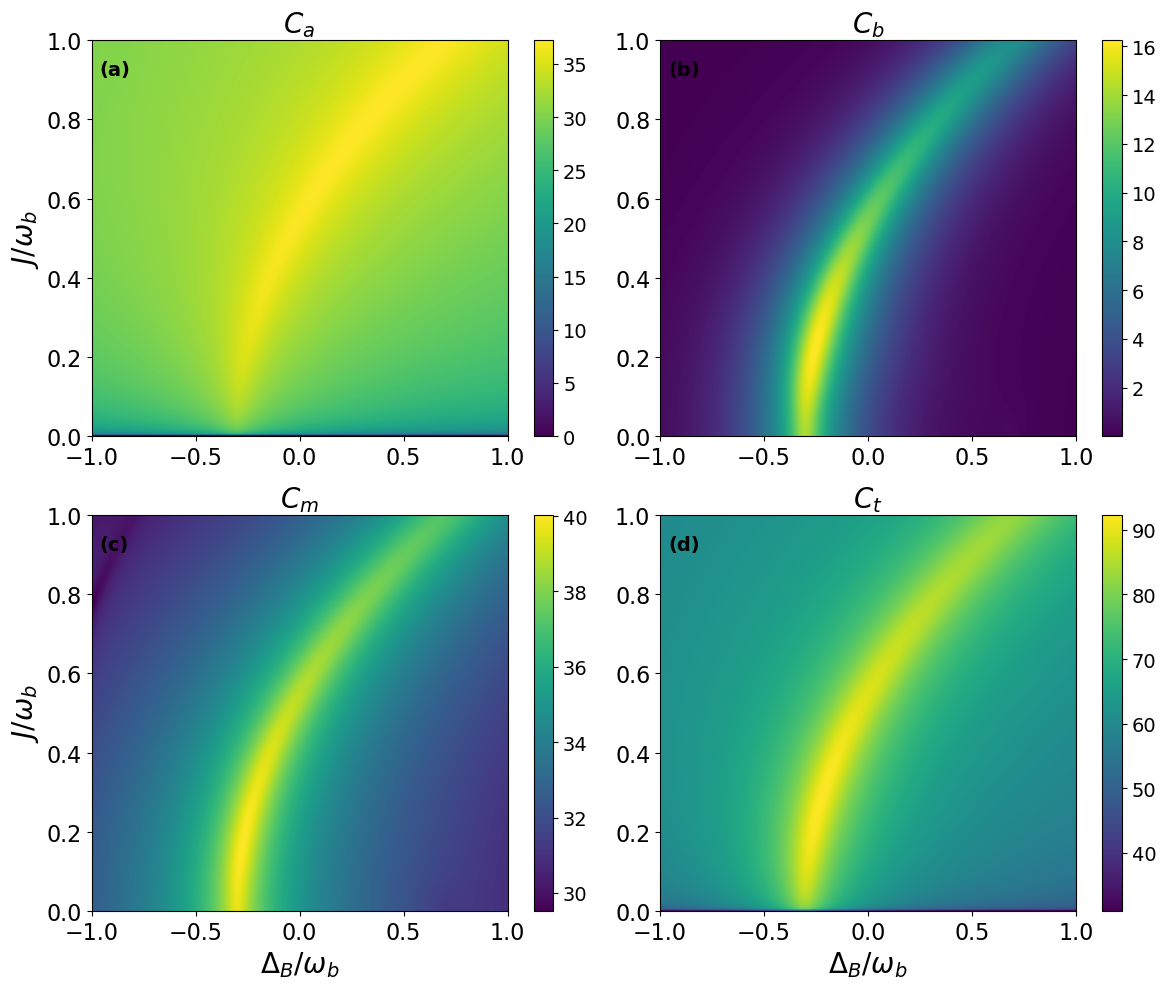}
    \centering
    \caption{Quantum coherences $C_a$,$C_b$,$C_m$,$C_t$ as functions of the $\Delta_B$ and $J$. The driving power is $P=6\times10^{-4}~\mathrm{mW}$, $\tilde\Delta_m=0.3\omega_b.$ And the other parameters are the same as those in Fig.\ref{fig2}.}
    \label{fig3}
\end{figure}

\begin{figure}[H]

    \includegraphics[width=\linewidth]{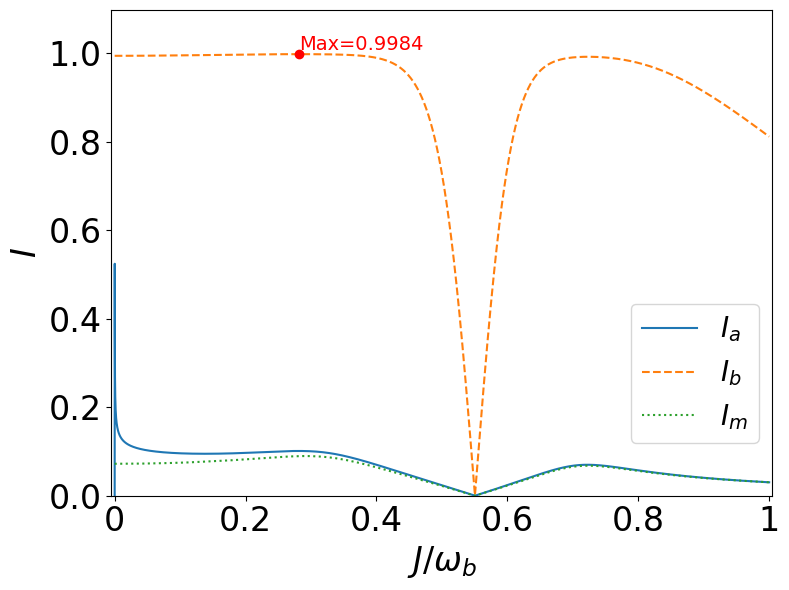}
    \centering
    \caption{Bidirectional contrast ration $I$ as functions of $J$. The driving power is $P=1\times10^{-5}~\mathrm{mW}$ and the angular frequency is $\Delta_B=0.2\omega_b$. And the other parameters are the same as those in Fig.\ref{fig3}.}
    \label{fig4}
\end{figure}

From Fig.\ref{fig3}(a), we find that when \( J \) is relatively small, the quantum coherence of the photonic mode remains essentially unchanged as \( \Delta_B \) varies. This behavior is expected because, in this regime, the photon mode is effectively decoupled from the magnon mode. Consequently, the nonreciprocal quantum coherence induced by the Barnett effect does not influence the coherence properties of the photon mode. We can see from Figure~4 that when \( J=0 \), \( I=0 \), indicating the absence of non-reciprocal quantum coherence. As \( J \) increases, the bidirectional contrast ratio exhibits a sudden jump followed by a gradual decrease. This behavior arises from the abrupt transition of the photon mode from decoupling with the magnon mode to weak coupling. We observe that quantum coherence is not a monotonic function of either the coupling rate \( J \) or the angular frequency \( \Delta_B \). Specifically, increasing \( J \) or \( \Delta_B \) does not necessarily lead to a higher  quantum coherence. Therefore, by appropriately tuning the coupling coefficient \( J \) and the angular frequency \( \Delta_B \), one can optimize the quantum coherence in experiments. Moreover, we find that at \( J = 0.54\omega_b \), the degree of nonreciprocity in quantum coherence across all modes vanishes. This phenomenon can be explained by starting from the steady-state solution. The steady-state value $m_s$ can be written in a simpler form under condition $\mid\tilde\Delta_m\mid,\mid\Delta_a\mid>>\kappa_a,\kappa_m$, that is, $m_s=\mathrm{i}\varepsilon_l\Delta_a/(J^2-\Delta_a(\tilde\Delta_m+\Delta_B))$. When \( \Delta_B = 0 \), meaning the YIG sphere remains stationary, no nonreciprocal steady-state solutions appear. At the condition$ J^2\approx \Delta_a\tilde\Delta_m$, i.e., $J \approx 0.54\omega_b$, the degree of nonreciprocity is minimized. As shown in Fig.\ref{fig4}, when the coupling parameter $J$ increases from 0 to 0.4, the nonreciprocity of the phonon mode remains close to $1$, with a peak value reaching 0.9984. This result demonstrates that the proposed scheme enables strong non-reciprocal behavior by utilizing the Barnett effect on a microscale platform. Such a mechanism provides a promising route toward on-chip integration of nonreciprocal phononic devices. 
%In addition, the feasibility of realizing non-trivial steady-state solutions based on the Barnett effect has been discussed in Ref.~[A], suggesting that experimental implementation of our scheme is within reach and warrants further investigation.

\begin{figure}[H]

    \includegraphics[width=\linewidth]{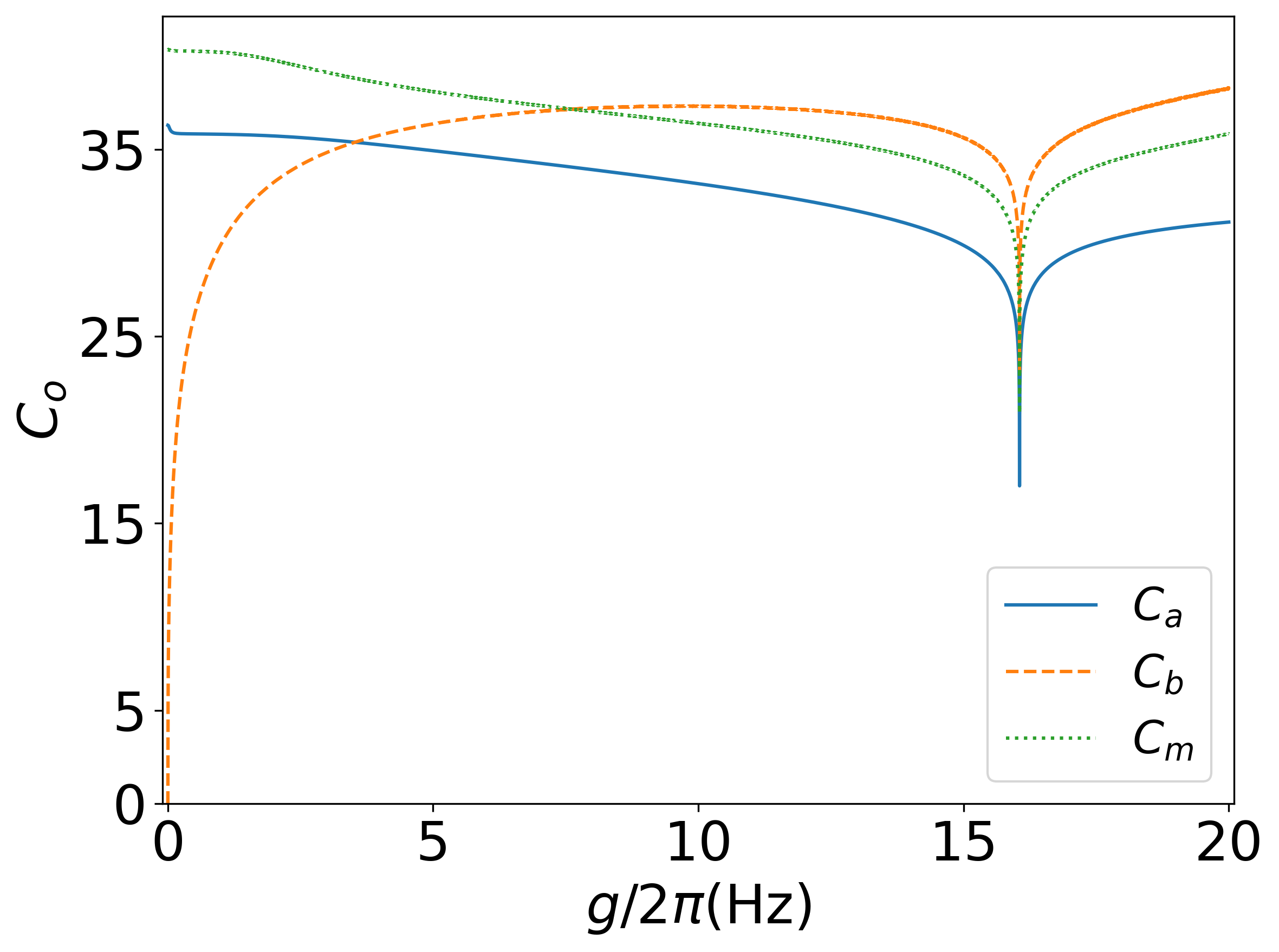}
    \centering
    \caption{Quantum coherences $C_a$,$C_b$,$C_m$ as functions of $g$.The driving power is $P=7\times10^{-4}~\mathrm{mW}$ and the magnon-photon coupling rate is $J=0.26\omega_b$. The angular frequency is $\Delta_B=-0.24\omega_b$. And the other parameters are the same as those in Fig.\ref{fig4}.}
    \label{fig5}
\end{figure}

As shown in Fig.\ref{fig5}, the quantum coherences $C_a$, $C_b$, and $C_m$ are plotted as functions of the magnon-phonon coupling rate $g$.
In addition, Fig.~2 shows that the quantum coherences of the individual modes $C_a$, $C_b$, and $C_m$ are not monotonic functions of the phonon-magnon coupling rate $g$. A dip appears at a certain critical coupling rate. This behavior can be understood as follows. As $g$ gradually increases, the effective magnomechanical coupling strengths $g_1$ and $g_2$ become larger. As a result, the quantum correlations in the system can be enhanced through the magnomechanical coupling. The vibration amplitude of the phonon mode also increases, which leads to an amplification of quantum fluctuations of the quadrature operators \(\delta X_O\) and \(\delta Y_O\).
In this case, the covariance matrix $V$ and the corresponding symplectic eigenvalues of the single mode $v_o$ change significantly, leading to a reduction in quantum coherence. In particular, when the coupling strength exceeds a certain threshold (e.g. $g > 8$~Hz), the coherence of the phonons $C_b$ can surpass that of the photon and magnon modes, i.e. $C_b > C_a, C_m$. These findings demonstrate that quantum coherence can be effectively redistributed from the photon and the magnon modes to the phonon mode, enabled by the direct magnon-phonon interaction and the indirect beam-splitter interaction between the photon and the phonon modes.

\begin{figure}[H]

    \includegraphics[width=\linewidth]{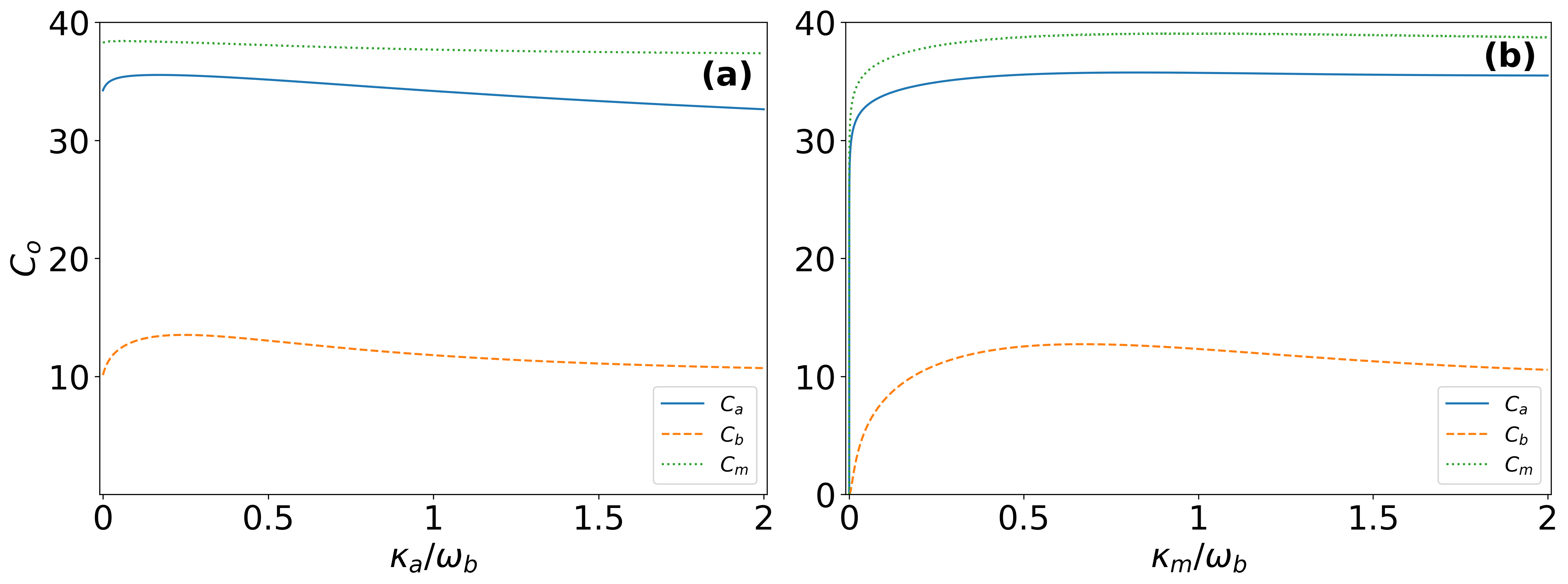}
    \centering
    \caption{Quantum coherences $C_a$,$C_b$,$C_m$ as functions of the  decay rates of the phonon mode and the magnon mode.The driving power is $P=1\times10^{-2}~\mathrm{mW}$ and the magnon-photon coupling rate is $J=0.4\omega_b$. The angular frequency is $\Delta_B=0.25\omega_b$. And the other parameters are the same as those in Fig.\ref{fig5}.}
    \label{fig6}
\end{figure}

In Fig.~\ref{fig6}, we analyze the impact of the cavity decay rates $\kappa_a$ and the magnon decay rate $\kappa_m$ on the quantum coherences. Interestingly, all the one-mode quantum coherences exhibit a non-monotonic behavior: they first increase and then decrease as the decay rates increase. However, the underlying physical mechanisms differ in each case.
In Fig.~\ref{fig6}(a), as $\kappa_a$ increases, the steady-state amplitudes of various modes are initially enhanced, leading to a growth in quantum coherences. However, further increase in $\kappa_a$ introduces stronger effective thermal noise, which dominates the dynamics and suppresses the one-mode quantum coherences.
In Fig.~\ref{fig6}(b), the optical decay $\kappa_m$ has two competing effects. On one hand, it depletes the cavity field, thereby reducing the quantum coherences. On the other hand, it contributes to the field buildup via the external drive, as described by the relation $\varepsilon_l = \sqrt{2\kappa_m P_l / \hbar \omega_l}$. When $\kappa_m$ is small, the driving effect dominates, enhancing the coherences. As $\kappa_m$ increases, the competition between optical decay and external pumping leads to a balance point where quantum coherences reach their maximum. Beyond this point, the detrimental effect of optical decay outweighs the benefit from the pump, resulting in a decline in the quantum coherences.

\section{\label{sec:level4}Conclusion:}
In conclusion, we theoretically proposed a scheme which generates the nonreciprocal quantum coherence in the CMM system via the Barnett effect. Our investigation shows that the nonreciprocity arises from the different system stability for applying the magnetic field along opposite directions. Moreover, with the selection of the proper parameters, nearly perfect nonreciprocity of quantum coherence can be achieved. The quantum coherence of the phonon mode can exceed that of the photon mode and the magnon mode because coherence is transferred between different modes. In addition, the quantum coherences are influenced by both cavity and the magnon decay rates, they all first increase and then decrease as the decay rates increase.
Our findings pave the way for the design of nonreciprocal quantum resources in the CMM systems and may find useful applications in achieving on-chip integration.

\section{\label{sec:level5}Acknowledgments:}

Mei Zhang thanks Junzhong Yang for helpful discussions. Jinhao Jia thanks Fan Zhou and Hejin Lv for helpful discussions. This work was supported by the National Natural Science Foundation of China under Grant No. 11475021 and the National Key Basic Research Program of China under Grant No. 2013CB922000.

\nocite{*}

\bibliographystyle{unsrt}

\bibliography{1}

\end{document}